\documentclass[twocolumn,preprintnumbers,amsmath,amssymb]{revtex4}

\usepackage{graphicx}
\usepackage{bm}

\begin{document}

\title{Cross-Kerr interaction in a four-level atomic system}

\author{Gary F. Sinclair and Natalia Korolkova}
 \address{School of Physics and Astronomy, University of St Andrews, North Haugh, St Andrews, KY16 9SS, Scotland}

\begin{abstract}
We derive the form of the cross-Kerr interaction in a four-level atomic system in the N-configuration.  We use time-independent perturbation theory to calculate the eigenenergies and eigenstates of the Schr\"{o}dinger equation for the system.  The system is considered as a perturbation of a Raman resonant three-level lambda scheme for which exact solutions are known.  We show that within the strong control field limit the cross-Kerr interaction can arise between two weak probe fields.  The strength of this nonlinear coupling is several orders of magnitude larger than that achievable using optical fibres.
\end{abstract}

\maketitle

\section{Introduction}
The cross-Kerr interaction involves controlling the refractive index experienced by one electromagnetic field by the intensity of another.  This nonlinear coupling between the field modes is vital in several quantum information protocols \cite{WNOQC, CTXPM, TEITKN} and has found applications in nondemolition photon measurement \cite{HEQNDD} and entanglement concentration \cite{PECCVSL, ECCVQS}.

\begin{figure}[t]
  \center
  \includegraphics[width=8cm]{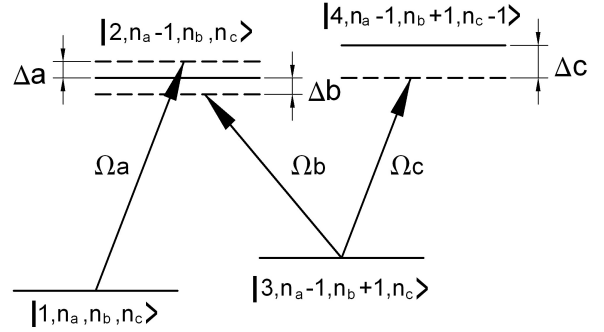}
  \caption{The four-level atomic system interacting with three electromagnetic field modes. $\Omega_x$ and $\Delta_x$ are the Rabi frequencies and single photon detunings corresponding to each electromagnectic field mode.}
  \label{int:fls}
\end{figure}

However, achieving a strong nonlinear interaction whilst avoiding self-phase modulation or large absorption has proved challenging.  Conventional methods involve either four-wave mixing in three-level atomic systems \cite{NOPEIT} or rely on the weak third-order susceptibility experienced in microstructured optical fibres \cite{NFO}.  It has been suggested by Schmidt and Imamo\v{g}lu \cite{GKNEIT} that a much stronger interaction is generated in the four-level atomic system (Fig. \ref{int:fls}).  The experimental feasibility of achieving a large cross-Kerr interaction for continuous-wave fields has been demonstrated \cite{OLKN}.  However it is only recently that a suitable method was suggested for group-velocity matched pulses \cite{LXPMSCP}, as might be necessary for the construction of quantum logic gates.

The Hamiltonian for a cross-Kerr interaction \cite{IQO} between two fields $\Omega_a$ and $\Omega_c$ is given by
\begin{equation}
  \hat{H}=\hbar K \hat{n}_a \hat{n}_c,
\end{equation}
where $K$ is the coupling strength and $\hat{n}_a$ and $\hat{n}_c$ are photon-number operators.  When acting on a state consisting of two modes, both of which are in number states ($\vert \psi(0) \rangle=\vert n_a \rangle \otimes \vert n_c \rangle$) the evolution exibits a cross-phase modulation of the form
\begin{equation}
  \vert \psi(t) \rangle = \exp(- i K n_a n_b t) \vert \psi(0) \rangle.
\end{equation}

We propose an intuitive and straightforward method to derive the form of the cross-Kerr interaction in the four-level atomic system in a N-configuration for continuous-wave fields.  By using time-independent non-degenerate perturbation theory \cite{QMFM} we find the eigenstates and eigenenergies of the time-independent Sch\"{o}dinger equation and show that given certain conditions one of these eigenstates gives rise to a cross-Kerr interaction.

The Hilbert space of the four-level atom plus three electromagnetic field mode system forms an infinite dimensional tensor product space that is spanned by the set of vectors
\begin{equation}
  \{ \vert i \rangle \otimes \vert n_a \rangle \otimes \vert n_b \rangle \otimes \vert n_c \rangle \}.
\end{equation}
Here $i \in [1, 4]$ denotes the atomic state and $n_x \in [0, \infty)$ describes the Fock state of the corresponding field mode $\Omega_x$.  It is assumed that the four-level atom plus three electromagnetic field modes form a closed system.   This enables us to decompose the entire space into four-dimensional subspaces that are invariant under the unitary evolution.  That is, given a system in a particular state, the evolution will remain in the subspace corresponding to absorption and creation of single photons due to transitions between the four atomic states \cite{ACPT}.  Omitting the tensor product symbols, the ordered basis is given by 
\begin{equation}
  \begin{array}{l}
    \{ \vert 1, n_a, n_b, n_c \rangle , \vert 2, n_a-1, n_b, n_c \rangle, \vert 3, n_a-1, n_b+1, n_c \rangle ,\\
    \vert 4, n_a-1, n_b+1, n_c-1 \rangle \}
  \end{array}
\end{equation}
Here the basis vectors are labeled by the numbers $n_x$, which give the number of photons in the field $\Omega_x$ when the atom is in the state $\vert 1 \rangle$.  Henceforth, we will further abbreviate the notation for the basis by refering only to the atomic state
\begin{equation}
  \{ \vert 1 \rangle , \vert 2 \rangle, \vert 3 \rangle, \vert 4 \rangle \}.
\end{equation}
We assume that the electromagnetic fields couple to the atomic energy levels by the electric-dipole interaction and that the rotating-wave approximation can be applied.  Working in an interaction picture we find that the Hamiltonian can be written in the bare atomic state basis as
\begin{equation}
  \hat{H} = 
  \hbar \left ( \begin{array}{cccc}
      0 & \Omega^*_a/2 & 0 & 0 \\
      \Omega_a/2 & \delta_1 & \Omega_b/2 & 0 \\
      0 & \Omega^*_b/2 & \delta_2 & \Omega^*_c/2 \\
      0 & 0 & \Omega_c/2 & \delta_3
    \end{array} \right ).
  \label{int:h}
\end{equation}
The form of the Hamiltonian is identical to that used in semi-classical calculations: the only difference being the definition of the Rabi-frequencies which are given by
\begin{eqnarray}
  \Omega_a &=& 2 g_a \sqrt{n_a},\\
  \Omega_b &=& 2 g_b \sqrt{n_b + 1},\\
  \Omega_c &=& 2 g_c \sqrt{n_c}.
\end{eqnarray}
Here, $n_x$ is the number of photons in the field mode $x$ when the atom is in the state $\vert 1 \rangle$ and $g_x$ is the coupling strength which is proportional to the electric-dipole matrix elements for each transition.  The multi-photon detunings are given by
\begin{eqnarray}
  \delta_1 &=& \Delta_a, \\
  \delta_2 &=& -\Delta_b + \Delta_a, \\
  \delta_3 &=& \Delta_c - \Delta_b + \Delta_a.
\end{eqnarray}
To simplify the application of perturbation theory we introduce a symmetry into lambda subsystem: we assume that the fields $\Omega_a$ and $\Omega_b$ are Raman-resonant with the two-photon transition from the state $\vert 1 \rangle$ to $\vert 3 \rangle$ ($\delta_2 = 0$).  This greatly simplifies the determination of the lambda subsystem dressed states.  In our Raman-resonant case the multi-photon detunings reduce to
\begin{equation}
  \delta_1 = \Delta_a, \qquad \delta_2 = 0, \qquad \delta_3 = \Delta_c.
\end{equation}
\section{Nondegenerate Time-Independent Perturbation Theory}
For the purpose of applying perturbation theory we split the system into two parts: the unperturbed system described by the Hamiltonian $\hat{H}_0$ and a weak perturbation described by $\hat{V}$.  The unperturbed system consists of the four atomic levels coupled by the two fields $\Omega_a$ and $\Omega_b$.  This constitutes a three-level lambda-scheme plus a fourth uncoupled level $\vert 4 \rangle$.

The weak perturbation $\hat{V}$ arises due to the third field $\Omega_c$ which couples the atomic levels $\vert 3 \rangle$ and $\vert 4 \rangle$.  We assume that the perturbation strength is dictated by a single multiplicitive parameter $\epsilon$ $(\epsilon= \Omega^0_c/2)$.  Therefore, the Hamiltonian can be written as
\begin{equation}
  \hat{H}=\hat{H}_0 + \epsilon \hat{V}.
\end{equation}
For the total Hamiltonian given above (\ref{int:h}) $\hat{H}_0$ and $\epsilon \hat{V}$ are
\begin{equation}
  \hat{H}_0 = 
  \hbar \left ( \begin{array}{cccc}
      0 & \Omega^*_a / 2 & 0 & 0 \\
      \Omega_a / 2 & \delta_1 & \Omega_b / 2 & 0 \\
      0 & \Omega^*_b / 2 & 0 & 0 \\
      0 & 0 & 0 & \delta_3
    \end{array} \right ),
\end{equation}
\begin{equation}
  \epsilon \hat{V} = 
  \frac{\hbar \Omega^0_c}{2}\left ( \begin{array}{cccc}
      0 & 0 & 0 & 0 \\
      0 & 0 & 0 & 0 \\
      0 & 0 & 0 & e^{-i \phi}\\
      0 & 0 & e^{i \phi} & 0
    \end{array} \right ).
\end{equation}
Before using perturbation theory, it is necessary to know the eigenstates of the unperturbed Hamiltonian $H_0$.  As mentioned previously, the eigenstates of the lambda-system have a particularly simple form for the special case of a Raman-resonance between the lower levels  \cite{PIO35}.  We choose the following normalised eigenbasis:
\begin{eqnarray}
  \vert \phi^{(0)}_1 \rangle &=& \frac{1}{G}( \Omega_b \vert 1 \rangle - \Omega_a \vert 3 \rangle ), \\
  \vert \phi^{(0)}_2 \rangle &=& \frac{1}{N_-}( \Omega^*_a \vert 1 \rangle + \Omega^*_b \vert 3 \rangle + 2 \lambda_{-} \vert 2 \rangle ),\\
  \vert \phi^{(0)}_3 \rangle &=& \frac{1}{N_+}( \Omega^*_a \vert 1 \rangle + \Omega^*_b \vert 3 \rangle + 2 \lambda_{+} \vert 2 \rangle ),\\
  \vert \phi^{(0)}_4 \rangle &=& \vert 4 \rangle.
\end{eqnarray}
The normalisation constants are given by
\begin{eqnarray}
  G^2 &=& \vert \Omega_a \vert^2 + \vert \Omega_b \vert^2, \label{ntipt:g} \\
  N^2_{\pm} &=& 2 \left ( G^2 + \delta_1^2 \pm \delta_1 \sqrt{\delta_1^2 + G^2} \right ),
\end{eqnarray}
and the corresponding eigenenergies ($\hat{H}_0 \vert \phi^{(0)}_n \rangle = E^{(0)}_n \vert \phi^{(0)}_n \rangle = \hbar \lambda^{(0)}_n \vert \phi^{(0)}_n \rangle$) are
\begin{eqnarray}
  \lambda^{(0)}_{1} &=& 0, \\
  \lambda^{(0)}_{2} &=& \lambda_- = \frac{1}{2} \left ( \delta_1 - \sqrt{\delta^2_1 + G^2} \right), \\
  \lambda^{(0)}_{3} &=& \lambda_+ = \frac{1}{2} \left ( \delta_1 + \sqrt{\delta^2_1 + G^2} \right ), \\
  \lambda^{(0)}_{4} &=& \delta_3.
\end{eqnarray}
Consider now the dynamical stability of these eigenstates.  Assuming that the system is weakly coupled to the environment, spontaneous emission will relax the system into its most radiatively stable eigenstate, but can then be neglected.  From inspection we can see that only the state $\vert \phi^{(0)}_1 \rangle$ is radiatively stable.  Moreover, this state also exibits no material polarisations between atomic levels associated with allowed electric-dipole transitions and is therefore non-interacting or {\it dark} to the applied fields.  Therefore, the system will relax from any initial mixed state into the {\it dark state} $\vert \phi^{(0)}_1 \rangle$, or its perturbed counterpart.

We now wish to calculate the approximate eigenenergies and eigenstates of the total Hamiltonian.  To find approximate solutions we assume that these can be expanded in a power series of the interaction strength $\epsilon$:
\begin{equation}
  E_n = \sum^{\infty}_{i=0} \epsilon^i E^{(i)}_n,
  \qquad
  \vert \phi_n \rangle = \sum^{\infty}_{i=0} \epsilon^i \vert \phi^{(i)}_n \rangle.
  \label{ntipt:exps}
\end{equation}
We can expand each term $\vert \phi^{(i)}_n \rangle$ of the power series in terms of these unperturbed Hamiltonian basis states $\vert \phi^{(0)}_n \rangle$:
\begin{equation}
  \vert \phi^{(i)}_n \rangle = \sum^{4}_{s=1} a^{s(i)}_n \vert \phi^{(0)}_s \rangle.
  \label{ntipt:exp}
\end{equation}
Substituting the expansions (\ref{ntipt:exps}) into the energy eigenvalue equation $\hat{H} \vert \phi_n \rangle = E_n \vert \phi_n \rangle$ one finds
\begin{equation}
  (\hat{H}_0 + \epsilon \hat{V}) \sum^{\infty}_{i=1} \epsilon^i \vert \phi^{(i)}_n \rangle = \sum^{\infty}_{j=0} \sum^{\infty}_{k=0} \epsilon^{j+k} E^{(j)}_n \vert \phi^{(k)}_n \rangle.
\end{equation}
We assume that this equation is true for terms involving each power of $\epsilon$ independently.  Using (\ref{ntipt:exp}) and the normalisation condition $\langle \phi_n \vert \phi_n \rangle = 1$, one finds a set of coupled equations for the $E^{(i)}_n$ and $a^{s (i)}_n$ terms.  These equations determine the eigenvalues and eigenvectors of the system \cite{QMFM}.

Let us first consider the perturbation caused to the eigenenergies.  In the eigenbasis of the unperturbed system the interaction $\hat{V}$ only has zero elements along the main diagonal.  Therefore we find that all the eigenenergies are unchanged to first order.  The first non-zero contributions are second order corrections given by
\begin{equation}
  E^{(2)}_n = \mathop{\sum^{4}_{s=1}}_{s \ne n} \frac{\vert \langle \phi^{(0)}_s \vert \hat{V} \vert \phi^{(0)}_n \rangle \vert^2}{E^{(0)}_n - E^{(0)}_s}.
\label{ntipt:e2}
\end{equation}
This results in the eigenenergy corrections ($E^{(i)}_n= \hbar \lambda^{(i)}_n$)
\begin{eqnarray}
  \lambda^{(2)}_1 &=& - \frac{\vert \Omega_a \vert^2}{\delta_3(\vert \Omega_a \vert^2 + \vert \Omega_b \vert^2)}, \label{ntipt:l12} \\
  \lambda^{(2)}_2 &=& - \frac{\vert \Omega_b \vert^2}{N^2_- (-2 \lambda_- + 2 \delta_3)}, \\
  \lambda^{(2)}_3 &=& \frac{\vert \Omega_b \vert^2}{N^2_+ (2 \lambda_+ + 2 \delta_3)}, \\ 
  \lambda^{(2)}_4 &=& \frac{4 (\delta_1 - \delta_3) \delta_3 + \vert \Omega_a \vert^2}{\delta_3 (G^2 + 4 \delta_3 (\delta_1 - \delta_3))}.
\end{eqnarray}
For example, the first energy eigenvalue is given to second order by
\begin{equation}
  E_1 \approx \hbar \lambda^{(0)}_1 + \frac{\vert \Omega_c \vert^2}{4} \hbar \lambda^{(2)}_1 = - \frac{\hbar \vert \Omega_a \vert^2 \vert \Omega_c \vert^2}{4 \delta_3 (\vert \Omega_a \vert^2 + \vert \Omega_b \vert^2)}. \label{ntipt:pe1}
\end{equation}
It is possible explain the form of this eigenvalue.  Namely, the energy correction arises due to the ac-Stark shift of the level $\vert 3 \rangle$ by the field $\Omega_c$.  Assuming that the system adiabatically follows this perturbation, then no atomic transitions between the unperturbed states will be induced.  The probability of the atom occupying the level $\vert 3 \rangle$ is $\vert \Omega_a \vert^2 /G^2$ if the system is in the state $\vert \phi^{(0)}_1 \rangle$.  Multiplying this by the ac-Stark shift produced by a weak field on that state, $-\vert \Omega_c \vert^2 / 4 \delta_3$, results in the energy shift
\begin{equation}
  \delta_E = \frac{\vert \Omega_a \vert^2}{G^2} \times -\frac{\vert \Omega_c \vert^2}{4 \delta_3}.
\end{equation}
Using (\ref{ntipt:g}), this simple calculation again yields the eigenvalue previously derived by the second-order perturbation theory (\ref{ntipt:pe1}).
\begin{figure}[t]
  \center
  \includegraphics[width=6cm]{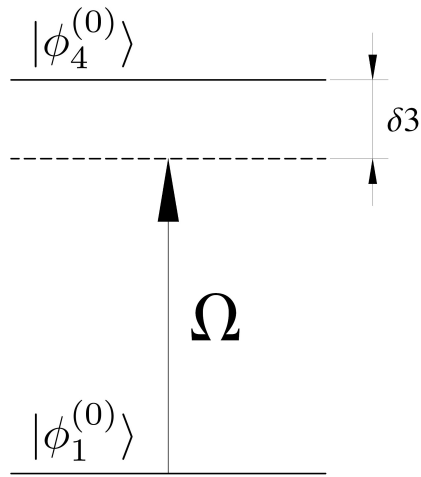}
  \caption{The two-level subsystem consisting of the composite field $\Omega = -\Omega_a \Omega_c / G$ coupling the ground state $\vert \phi^{(0)}_1\rangle$ to the excited state $\vert \phi^{(0)}_4 \rangle$.}
  \label{ntipt:tls}
\end{figure}

It should be noted that the eigenvalue correction (\ref{ntipt:l12}) fails to provide a good approximation when the detuning $\delta_3$ tends towards zero.  However, in the summation (\ref{ntipt:e2}) the only term contributing to the energy $E^{(2)}_1$ of the dark state $\vert \phi^{(0)}_1 \rangle$ arises due to the coupling of this state to the excited state $\vert \phi^{(0)}_4 \rangle$.  Therefore, to second order in the perturbing field $\Omega_c$, the dark-state $\vert \phi^{(0)}_1 \rangle$ behaves as if it was the lower level in a two-level subsystem (Fig. \ref{ntipt:tls}).

We may introduce the ordered basis for the pseudo-two-level subsystem
\begin{equation}
  \vert \phi^{(0)}_1 \rangle = \vert 1 \rangle = \left [ \begin{array}{c} 1 \\ 0 \end{array} \right ], \qquad \vert \phi^{(0)}_4 \rangle = \vert 2 \rangle = \left [ \begin{array}{c} 0 \\ 1 \end{array} \right ].
\end{equation}
The matrix elements of this Hamiltonian are given by $[H_{TLS}]_{mn} = \langle m \vert H_0 + \epsilon V \vert n \rangle$.  This gives
\begin{equation}
  H_{TLS} = \hbar \left ( \begin{array}{cc}
      0 & \Omega^*/2 \\
      \Omega/2 & \delta_3
    \end{array} \right ),
\end{equation}
where we define the composite field $\Omega=-\Omega_a \Omega_c / G$.  For a two-level system the lower eigenenergy is given by
\begin{equation}
  E = \frac{\hbar \delta_3}{2} \left ( 1 - \sqrt{1 + \frac{\vert \Omega_a \vert^2 \vert \Omega_c \vert^2}{\delta^2_3 G^2}} \right ).
\end{equation}
This is valid up to second order in $\Omega_c$, but unlike the non-degenerate perturbation theory result this correction also holds for all values of detuning $\delta_3$.  In the limit of strong detuning we again recover the standard perturbative result dervied previously (\ref{ntipt:pe1}).

\section{Cross-Kerr Interaction}
We now wish to show that when the system has relaxed into the perturbed dark state, a cross-Kerr interaction will arise between the fields $\Omega_a$ and $\Omega_c$.  The perturbed dark-state is given to first order by
\begin{equation}
  \vert \phi_1 \rangle \approx \frac{1}{G} \left ( \Omega_b \vert 1 \rangle - \Omega_a \vert 3 \rangle + \frac{\Omega_a \Omega_c}{2 \delta_3 G} \vert 4 \rangle \right ).
\end{equation}
Essentially, by calculating the approximate eigenenergy of the perturbed dark state $\vert \phi_1 \rangle$ we have determined the evolution of the system when relaxed into this eigenstate.  This evolution is given approximately by
\begin{equation}
  \vert \psi(t) \rangle \approx \exp (-i E^{(2)}_1 t / \hbar ) \vert \psi(0) \rangle,
\end{equation}
where the intial state is $\vert \psi(0) \rangle = \vert \phi_1 \rangle$.

To shown that this evolution gives rise to a cross-Kerr interaction we place one further limitation upon the system.  We assume, that the electromagnetic field $\Omega_b$ is much stronger than the fields $\Omega_a$ and $\Omega_c$ ($\vert \Omega_b \vert \gg \vert \Omega_a \vert, \vert \Omega_c \vert$).  On applying this condition the eigenenergy and eigenstate simplify to
\begin{eqnarray}
   E_1 &\approx& - \frac{ \hbar \vert \Omega_a \vert^2 \vert \Omega_c \vert^2}{4 \delta_3 \vert \Omega_b \vert^2}, \\
   \vert \phi_1 \rangle &\approx& \vert 1 \rangle \otimes \vert n_a \rangle \otimes \vert  n_b \rangle \otimes \vert n_c \rangle.
\end{eqnarray}
Recalling again the definition of the Rabi frequencies and subsituting these into the energy eigenvalue, we find
\begin{equation}
   E_1 \approx - \frac{ \hbar \vert g_a \vert^2 \vert g_c \vert^2 n_a n_c}{\delta_3 \vert g_b \vert^2 (n_b+1)}.
\end{equation}
Since the eigenstate of the system is now approximately equal to the bare state $\vert 1, n_a, n_b, n_c \rangle$ we can replace the numbers $n_a$, $n_b$ and $n_c$ in the eigenvalue $E_1$ with the corresponding number operators $\hat{n}_a$, $\hat{n}_b$ and $\hat{n}_c$.  This is possible since the state vector $\vert 1, n_a, n_b, n_c \rangle$ is an eigenstate of the operators $\hat{n}_x$ with eignevalues $n_x$.

The evolution of the system is therefore described by
\begin{equation}
  \vert \psi(t) \rangle = \exp (- i K \hat{n}_a \hat{n}_c t) \vert \psi(0) \rangle, 
\end{equation}
where we have defined the strength of the cross-Kerr non-linearity as
\begin{equation}
  K = - \frac{ \vert g_a \vert^2 \vert g_c \vert^2}{\delta_3 \vert g_b \vert^2 (\hat{n}_b+1)}.
\end{equation}
Importantly, the field $\Omega_b$ only modulates the strength of the cross-Kerr effect, although the interaction stregth is limited by the requirement that $\vert \Omega_b \vert \gg \vert \Omega_a \vert, \vert \Omega_c \vert$.

\section{Limitations}
It has been assumed throughout that a Raman-resonance exists between the lower atomic levels.  This condition is necessary for the generation of a pure cross-Kerr nonlinearity.  Moreover, without this constraint the system would exibit an additional phase delay and self-phase modulation of the field $\Omega_a$.  Due to the narrow bandwidth of the EIT window, we expect the system to be very sensitive to small changes in two-photon detuning.  For instance, introducing a time-dependence to the field $\Omega_a$ would invalidate the Raman-resonance condition due to the spectral width of the pulse.  

However, since the field $\Omega_c$ need only be strongly detuned, increasing the spectral width should not have a detrimental effect.   Each spectral component would experience an independent cross-Kerr interaction.  Nonetheless, only the field $\Omega_a$ experiences slow-light conditions so the interaction time of probe pulses is constrained by the group-velocity mismatch.  Extending this analysis to the non-Raman-resonant regime will be the subject of a further publication.

\section{Conclusion}
By applying non-degenerate perturbation theory to the time-independent Schr\"{o}dinger equation we have shown that the cross-Kerr interaction can arise in an atomic four-level system in the N-configuration.  In paticular, we perturb a Raman-resonant lambda system by introducing a fourth atomic level weakly coupled to one of the lambda-system ground states.  The system relaxes into the perturbed dark-state of the lambda subsystem and gives rise to the cross-Kerr nonlinearity in the strong pump ($\Omega_b$ ($\vert \Omega_b \vert \gg \vert \Omega_a \vert, \vert \Omega_c \vert$) and strong detuning limits ($\delta_3 \gg \Omega_c$).  Importantly, this interaction has been shown theoretically \cite{GKNEIT} and experimentally \cite{OLKN} to be several orders of magnitude larger than that produced by other contemporary methods.

\bibliography{references}
\bibliographystyle{unsrt}

\end{document}